# Microwave Vortex-Beam Emitter Based on Spoof Surface Plasmon Polaritons

*Jia Yuan Yin[1,2], Jian Ren[3], Lei Zhang[1,2], Haijiang Li[4], and Tie Jun Cui[1,2]\**

*Corresponding Author: E-mail: tjcui@seu.edu.cn

[1]State Key Laboratory of Millimeter Waves, School of Information Science and Engineering, Southeast University, Nanjing 210096, China
[2]Synergetic Innovation Center of Wireless Communication Technology, Southeast University, Nanjing, 210096, China
[3]State Key Laboratory of Millimeter Waves and Department of Electronic Engineering, City University of Hong Kong, Kowloon, Hong Kong SAR, China
[4]Jiangsu Xuantu Technology Co., Ltd., 12 Mozhou East Road, Nanjing 211111, China

Since the orbital angular momentum (OAM) being investigated intensively in the optical region, there are growing interests in employing OAM to solve the problem in wireless communications as a new method. It is found that the independence between different OAM modes is crucial to wireless communications. Motivated by the tremendous potential of OAM in communication systems, we propose a novel method to generate vortex beams by spoof surface plasmon polaritons (SPPs). A looped double-layer spoof SPP waveguide is applied to realize the transmission of electromagnetic waves. Beam emitting is accomplished through a series of circular patches, whose role is not only the radiation units but also resonators giving rise to the phase shifts required by the vortex beam. The proposed method is validated by both numerical calculation and experiment. The measured results show that the spoof SPPs are radiated by the circular patches and the vortex beam carrying different OAM modes are observed at different frequencies. This is the first time to generate the OAM modes by spoof SPPs. The proposed method possesses smaller size and is much easier to be integrated into integrated circuits. The simple structure and design procedure make the proposed method promising in future wireless communication systems.

# 1. Introduction

Optical vortices are light beams with helical phase fronts and azimuthal components of the wave vectors. Since the fact that photons in optical vortices carry orbital angular momentum (OAM) was discovered, [1] OAM modes have been investigated intensively in the field of optical microscopy, [2] micromanipulation, [3-7] super-resolution imaging, [8,9] and quantum information technologies. [10-15] After a series of profound studies, there are growing interests in employing OAM to solve the problem in wireless communication as a new method. The first radio OAM mode (or twisted wave) simulation was performed in 2007, providing a theoretical foundation for OAM-based wireless communications. [16] In 2012, the first experiment of the wireless radio transmission applying the vortex beam was completed, which verifies that the OAM-carried electromagnetic (EM) waves make contributions to the increase of communication capacity without increasing the bandwidth. [17] It has also been demonstrated that the EM waves with different OAM modes can work independently to each other in the wireless communications. That is to say, if the mode of the transmitting OAM wave is designed as l, spiral phase plates that generate –l mode OAM should be applied to receive the Gaussian beam by the receiving antenna. [18]

Owing to the novel advantages of OAM modes, approaches to generate electromagnetic (EM) waves carrying OAMs become a focused research area. Up to now, there have been many methods to generate OAMs. The most attractive way among them is to use spiral phase plates due to its simple structure and practicality. [5,19-22] This method has been widely adopted in optics, followed by being introduced to the microwave and millimeter wave bands. The spiral phase plate is such a diffractive element that can modulate the phase of EM fields by its optical thickness. The other widely used method is antenna array. [23-27] Although the theory and technology foundation have already almost been perfect, the complex phase-shifting network is necessary to generate the required rotation phase. Not only the phase relationship among the radiation units should be guaranteed, but also the power needs to be consistent for the quality of OAMs. When the mode number of OAM is larger, more antennas are needed, which will result in significant increase of complexity and cost.

Recent progresses in metasurface offer new ways of manipulating phase distributions of EM waves. [28-32] By properly arranging the subwavelength unit cells with different dimensions on the metasurface, the EM waves interacted with the metasurface can be translated into the one with the OAM mode, and the total phase range of the metasurface determines the mode number of OAMs. Meanwhile, more and more attention has been paid on single cavities due to the simple structures and easy fabrication, such as whispering-gallery mode resonators. [33-35] The structures supporting whispering-gallery mode have interactions with the radiation mode, leading to the appearance of OAM modes. Half-mode substrate integrated waveguide antenna was also demonstrated to have the ability to generate vortex beams. [36] There are advantages of these existing methods but also disadvantages, such as the complexity of the structure, the difficulty in controlling the OAM mode, and the limited number of OAM modes.

In this article, a novel method is proposed to generate the vortex beams, which is based on the spoof surface plasmon polaritons (SPPs). A looped double-layer spoof SPP waveguide is applied to construct the transmission route of the EM waves, while a series of circular patches is set beside the spoof SPP waveguide for beam emitting. At the same time, the circular patches also function as resonators modulating the phase of the radiation beam. The theoretical calculation of the beam emission is illustrated, while the prototype is simulated and measured. The measured results agree very well with the simulations, which are also predicted by the theoretical analysis. We show that the vortex beams with different OAM modes are generated at different frequencies without any changes in the structure. Moreover, the proposed structure can be added into the integrated circuits easily as an important part of the wireless communications. To the best of our knowledge, this is the first time to produce OAM modes using spoof SPPs. The proposed method not only expands the application area of spoof SPPs, but also provides a much easier way to generate vortex beams.

## 2. Results
## 2.1. Structure and Operating Principle

Since the vortex beam owns helical phase fronts and azimuthal component of wave vector, the key to generate the vortex beam is the rotated phase distribution and radiation. **Figure 1(a)** illustrates a prototype of the proposed structure, in which a single-side corrugated metallic strip is used to form the spoof SPP waveguide. Because of the loop structure, the spoof SPP waveguide is designed on both sides of the dielectric substrate in order to avoid the overlap. The top and bottom SPP waveguides are connected by a metallic via, as shown in **Figure 1(b)**. It was declared that the metallic via can form surface mode on the transition part, instead of the waveguide mode caused by the boundaries of the two layers. [37] This results in the high transmission efficiency, especially when the thickness of the substrate is large. The substrate in this particular design is chosen as commercial printed circuit board, F4B, with a relative permittivity of 2.65 and loss tangent of 0.003. To reduce the effect coming from the overlap between the top and bottom SPP waveguides, the thickness of the substrate is set as 3mm. The spoof SPP waveguide is made to be a loop with radius of 80 mm, for the purpose to form the whole length as an integer times of wavelength at the expected radiation central frequency (6 GHz). The radius of the circular patch is set as 8 mm for the same sake, so that the single-pass phase shift after a circular resonator can achieve $2\pi$. The reason of this set will be illustrated in the following. The pictures of the proposed design are also given in **Figure 1(c)**.

Consider the spoof SPP waveguide composed of the unit cell shown in **Figure 2**, whose detailed size is designed as p=5 mm, w=5 mm, a=2 mm, and h=4 mm. The dispersion curve is obtained through numerical method. From the dispersion curve, it can be seen obviously that the spoof SPP waveguide owns different propagating constants (k) at different frequencies. That is to say, when the EM waves propagate along the spoof SPP waveguide for the same physical distance, the phase shift is different at different frequencies. If the spoof SPP waveguide is made to be a loop, the rotated phase distribution will be achieved along the spoof SPP waveguide and a phase gradient along the azimuth direction will be obtained. These issues are the solid foundation to the OAM modes.

As the phase gradient along the azimuth direction can be realized through the variant of spoof SPP waveguide, another key issue is the efficient radiation of EM waves to free space. Here, a series of circular metal patches are used as radiators. It has been demonstrated that when the circular patches are placed around the spoof SPP waveguide, the SPP waves are easily coupled to the patches for efficient radiations. [38] Resonator is the other role of such circular metal patches in the next place. When the circular patches are put beside the spoof SPP waveguide, a phase lag appears along with the energy coupling. To illustrate this issue, we consider a single circular metal patch placed beside the SPP waveguide, as shown in **Figure 3(a)**. The relationship among the input electric field $E_1$, the output electric field $E_2$, and the circulatory electric fields $E_3$ and $E_4$ can be established according to the coupling relationship and the electric field in the feedback route:

$$\begin{pmatrix} E_2 \\ E_4 \end{pmatrix} = \begin{pmatrix} \rho & i\kappa \\ i\kappa & \rho \end{pmatrix} \times \begin{pmatrix} E_1 \\ E_3 \end{pmatrix} \tag{1}$$

where $\rho$ and $\kappa$ are the self-coupling coefficient and mutual-coupling coefficient, respectively. Both factors are supposed to be independent of frequency with satisfying $\rho^2+\kappa^2=1$. Also the circulatory electric field $E_4$ becomes $E_3$ after the propagation through the feedback route, as depicted as:

$$E_3=e^{-0.5\alpha L}e^{i\omega\tau}E_4=ae^{i\varphi}E_4 \tag{2}$$

where $\alpha$ is the attenuation of the circular resonator, $\varphi$ denotes the single-pass phase shift, $a$ is the single-pass amplitude transfer factor, $\tau$ signifies the single-pass transition time, and $\omega$ is the frequency. From Equation (1) and (2), the relationship between the input electric field E1 and output electric field $E_2$ can be obtained as:

$$t = \frac{E_2}{E_1} = \frac{\rho - ae^{i\varphi}}{1 - \rho ae^{i\varphi}} \tag{3}$$

Then the phase lag introduced by the circular resonator is determined by $\Phi=arg$ (*t*). Therefore, the final radiation phase is a combination of the phase from spoof SPP waveguide and that brought about from the circular metal patches. As a consequence, if the total phase shift around the loop is *l* times of 2π, then the mode number of OAM is *l*.

In order to get an intuitive recognition of the phase lag introduced by the circular patches, we suppose that the single-pass amplitude transfer factor is 1. Then the effective phase lags after the circular resonator under different self-coupling coefficients are described in **Figure 3(b)**. We note that, no matter what the self-coupling coefficient is, the effective phase-shift difference between the maximum single-pass phase shift and the minimum single-pass phase shift keeps 2π all the time. That is to say, if the single-pass phase shift after the circular resonator is made to be 2π, then the effective phase-shift difference between the situations with and without the circular resonator achieves 2π. This conclusion is necessary to the design of the OAM structures.

## 2.2. Analysis and Theoretical Arithmetic

To analyze the OAM modes carried by the radiation beam, the radiation at the central frequency is taken as an example. From the dispersion curve given in Figure 2, the propagating constant at 6 GHz is calculated as 194.7, corresponding to the waveguide wavelength of 32 mm approximately. Obviously, the circumference of the spoof SPP waveguide loop is about 15 times of the waveguide wavelength, indicating that the phase shift around the loop is 15 times of $2\pi$ if there are no circular patches added. When the circular patches are located beside the spoof SPP waveguide, the phase shift will have a significant change. From the analysis above based on Equation (3) and Figure 2(b), it is known that once there is a circular patch, there will be a phase lag of $2\pi$. In this particular design, there are 15 patches in total, resulting in a total phase lag of 15 times of $2\pi$. Combined with the phase provided by the spoof SPP waveguide, we conclude that the mode number ($l$) of the radiation beam around 6 GHz is zero. If the OAM mode number ($l$) is expected to be 1, the phase provided by the spoof SPP waveguide should be 16 times of $2\pi$, and then the total phase shift can achieve $2\pi$. According to the relationship between the propagating constant and phase shift, it is found that this condition is satisfied around 6.3 GHz. In a similar way, $l = 2$ appears at about 6.6 GHz. Furthermore, when the phase provided by the spoof SPP waveguide is less than the phase lag introduced by the circular metal patches, a reversal in the phase rotation turns up. At about 5.8 GHz, the phase provided by the spoof SPP waveguide is 14 times of $2\pi$, leading to the OAM mode number $l = -1$. Similarly, the situation with $l = -2$ can be reached at around 5.5 GHz. Table I provides the precise values of frequencies corresponding to the relevant OAM modes. From the comparison between the theoretical arithmetic and full-wave simulation, the accuracy of the analysis can be corroborated since the deviation of frequency is tiny. The cause of these subtle errors can be blamed to the approximate evaluation of the propagation constant, which is inevitable.

The far field radiation patterns at the predicted frequencies are also calculated through MATLAB, serving as a preliminary evidence of the proposed principle and design. The calculated results are given in Supporting Information.

### 2.3. Simulations and Measurements

To illustrate the efficiency of the radiation, the proposed structure is simulated and fabricated for measurements. All materials in the experiments are the same as those in the simulations. Scattering (*S*) parameters (i.e., reflection coefficients $S_{11}$ and transmission coefficients $S_{21}$) are measured through the Agilent Vector Network Analyzer (VNA), as shown in **Figure 4**. With reference to Figure 4(a), we observe that the reflection coefficients are less than -10dB during the whole radiation frequency range, indicating the good impedance match. From Figure 4(b), we clearly see the resonant frequencies corresponding to the relevant OAM modes. The good agreements between the simulated and measured results, as well as the comparison in Table I, firmly validate the theory mentioned before. Combined the two kinds of coefficients, we can conclude that the radiation efficiency is fairly high. Most of the energy is radiated to the free space by the proposed structure. Owing to the limitation of the experiment system, the radiation efficiency cannot be measured. But we give the simulated efficiency, as well as the gain of the proposed structure at corresponding frequencies in Supporting Information.

In order to further verify the above analysis, a near-field experiment was performed, as shown in **Figure 5**. The comparison between the simulated and measured near-field distributions is presented in **Figure 6**, in which Figure 6(a)-(e) show the simulated amplitudes, Figure 6(f)-(j) are the measured ones, while Figure 6(k)-(o) give the simulated phase distributions, and Figure 6(p)-(t) are the measured ones. The commercial software, CST Microwave Studio, was used to simulate the designed structure. Because of the limitation of the computer resources, the observation plane is set at 500 mm (10 times of the wavelength at the central frequency) above the proposed structure with an area of 370×370 mm$^2$. The near-field measurements were completed in an anechoic chamber by using near-field antenna test system, which is composed of a fixed platform and a position controllable probe connected to the Agilent VNA. One of the two 50Ω-coaxial cables acts as the connection between VNA and the sample. The other connects the probe and VNA. The vertical probe located in front of the sample with a distance of 600 mm (12 times of the wavelength at the central frequency) as the receiver. During the measurement, the probe moves along the x- and y- directions step by step, and the phase distributions are plotted by the measured data via MATLAB. The whole observation plane in the measurement is set as 370×370 mm$^2$, covering the main radiation region. The good agreement can be observed obviously from the comparison with full-wave simulations. With reference to the figures, the mode number of OAM beam is conspicuous. We can get $l$ = -2, -1, 0, 1, 2 at 5.5, 5.8, 6.0, 6.3, and 6.6 GHz, respectively, corresponding to the forecasts from the analysis.

From the simulated and measured amplitude distributions of the proposed structure, we can conceive the far-field radiation pattern. However, to get the precise cognition, the far-field radiation patterns are also measured for further proof of the vortex-beam generation (see **Figure 7**). Figure 7(a)-(e) and 7(f)-(j) illustrate the simulated and measured three-dimensional (3D) radiation patterns, respectively. At 6 GHz, an ordinary beam is obtained since the mode number is zero, as has been predicted before. Hollow radiation patterns are achieved at other frequencies as expected, except for some imperfect amplitude performance. The main reason for this issue owns to the continuous leak of the EM waves, resulting in the unequal radiation amplitudes from different metal patches. It is worth noting that the simulated and measured results keep almost the same, but still have some delicate differences. This is owing to the different scale in the two kinds of results. In the experiment, the 3D radiation patterns are created through the measurement system directly, and the scale cannot be adjusted. But the scales of the simulated results are modified for a clearer cognition of the radiation situations. Both simulated and measured results prove that the previous analysis is accurate and the proposed method to generate vortex beams by spoof SPPs is effective.

## 3. Conclusion

In summary, we proposed a novel and simple method to generate vortex beams by spoof SPPs. The transmission route of the EM waves was formed by a looped double-layer spoof SPP waveguide. The beam emitting was realized through a series of circular metal patches. Such circular patches are used not only for the radiation units, but also for resonators to produce the phase shift. The proposed method was verified by both numerical calculation and experiments, and the measurement results agree very well with the numerical simulations as well as the theoretical predictions. The vortex beams with different OAM modes were observed at different frequencies. The mode numbers of $l$ =-2, -1, 0, 1, 2 appear at 5.5, 5.8, 6.0, 6.3, and 6.6 GHz, respectively. This is the first time to generate OAM modes by spoof SPPs. The simple structure and design procedure make the proposed method promising in future wireless communication systems. Compared with previous methods, this smaller structure possesses more flexible OAM modes, and is much easier to be integrated into the integrated circuit. This work is not only an expansion to the applications of spoof SPPs, but also a simplifier of the vortex-beam generation.

**Supporting Information**
Figure S1. Calculated radiation patterns at different frequencies through MATLAB.
Table S1. Simulated gain and efficiency of the proposed structure at relevant frequencies.

**Acknowledgements**  This work was supported in part from the National Natural Science Foundation of China (61631007, 61571117, 61302018, 61401089, 61501112, 61501117), and in part from the 111 Project (111-2-05).

**Keywords**: spoof surface plasmon polaritons, vortex beam, orbital angular momentum

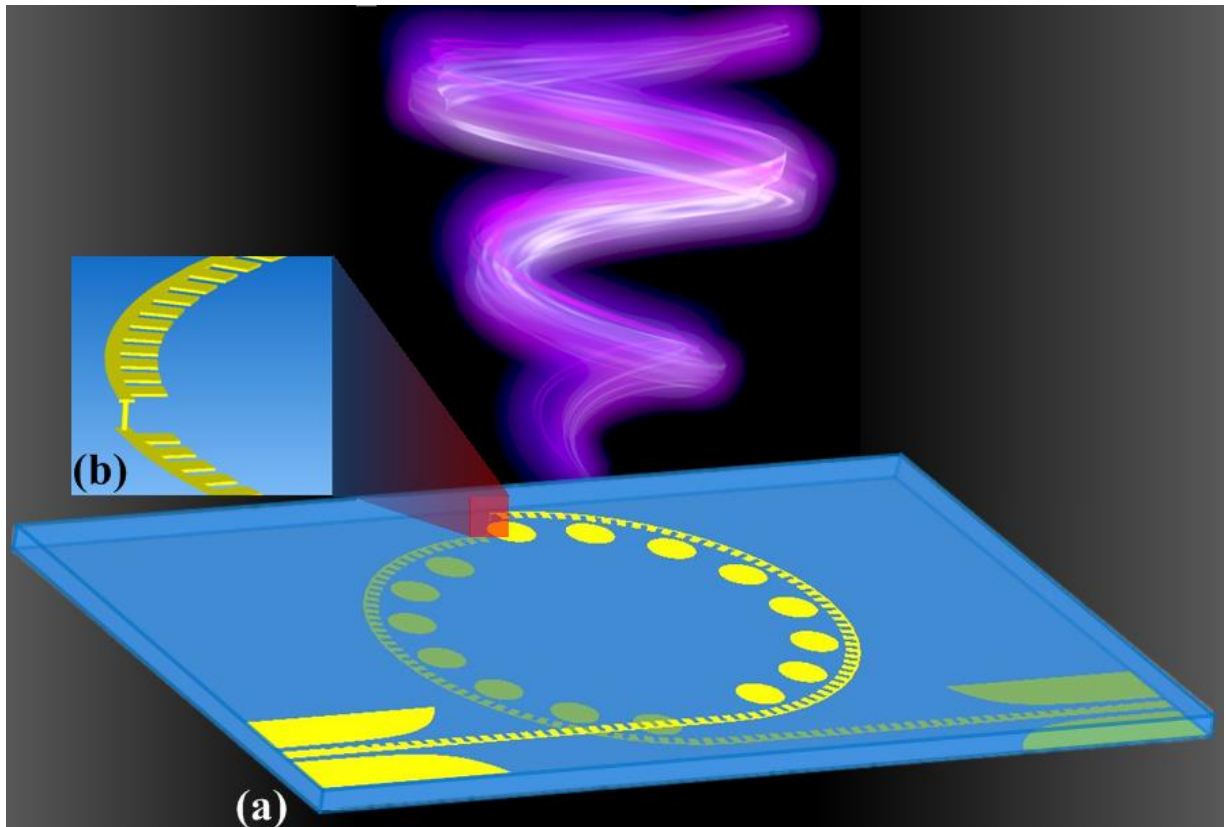

**Figure 1.** (a) Prototype of the vortex beam emitter, in which the yellow part is modeled as copper in the simulation and the blue part is the dielectric substrate. Particularly, the lighter yellow denotes the structure on the top of substrate, while the darker yellow represents the structure on the bottom of substrate. (b) Detailed illustration of the connection between the top and bottom structures.

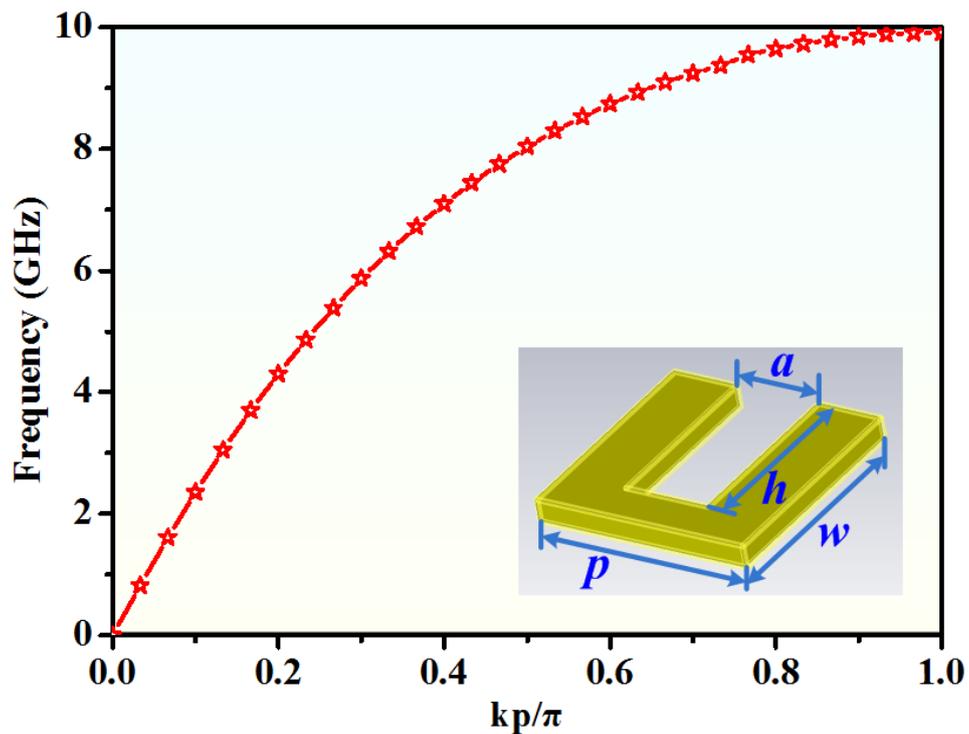

**Figure 2.** Dispersion curve of the unit cell form the spoof SPP waveguide.

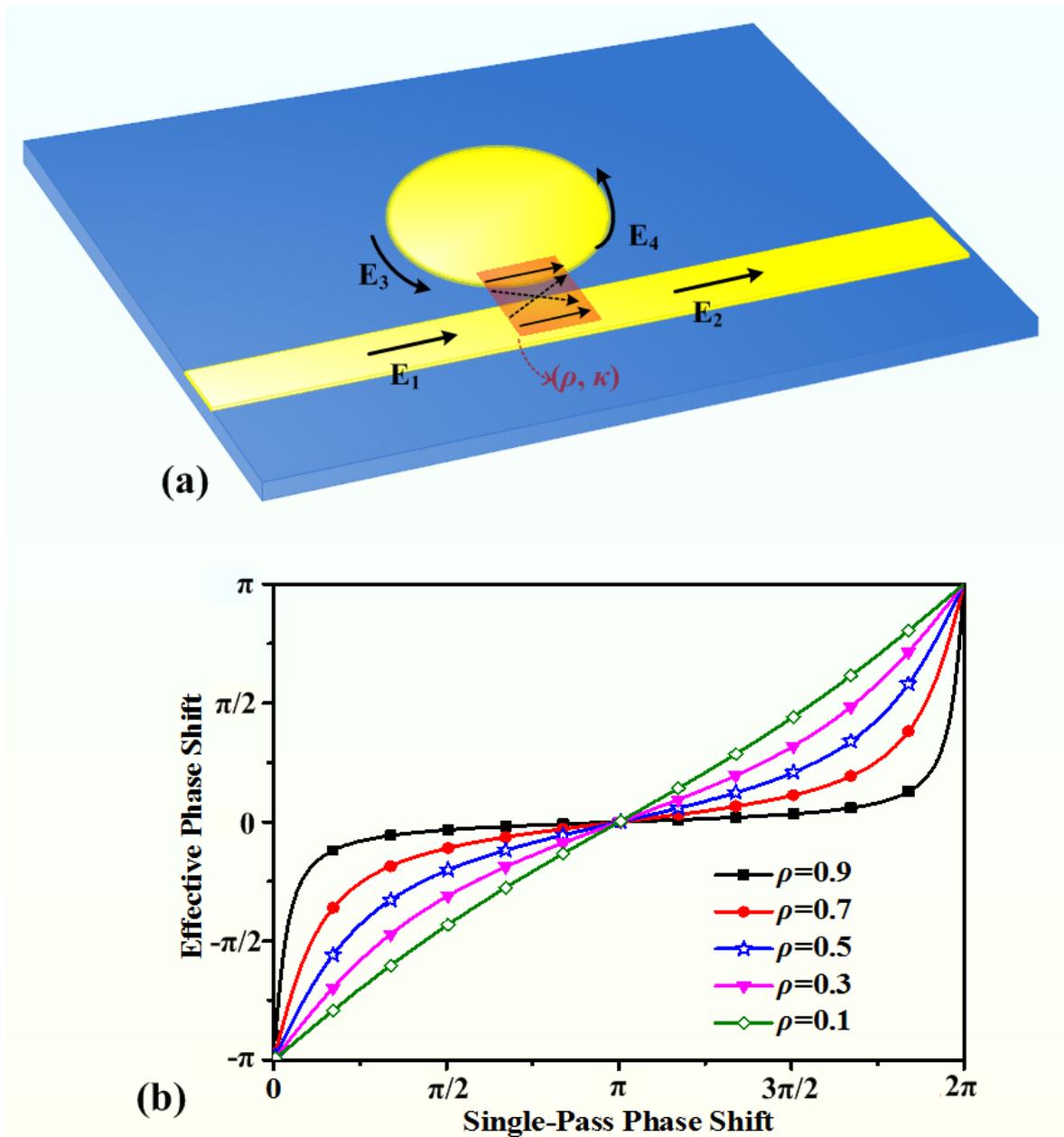

**Figure 3.** (a) Schematic diagram of a single circular resonator. (b) Effective phase shifts under different self-coupling coefficients.

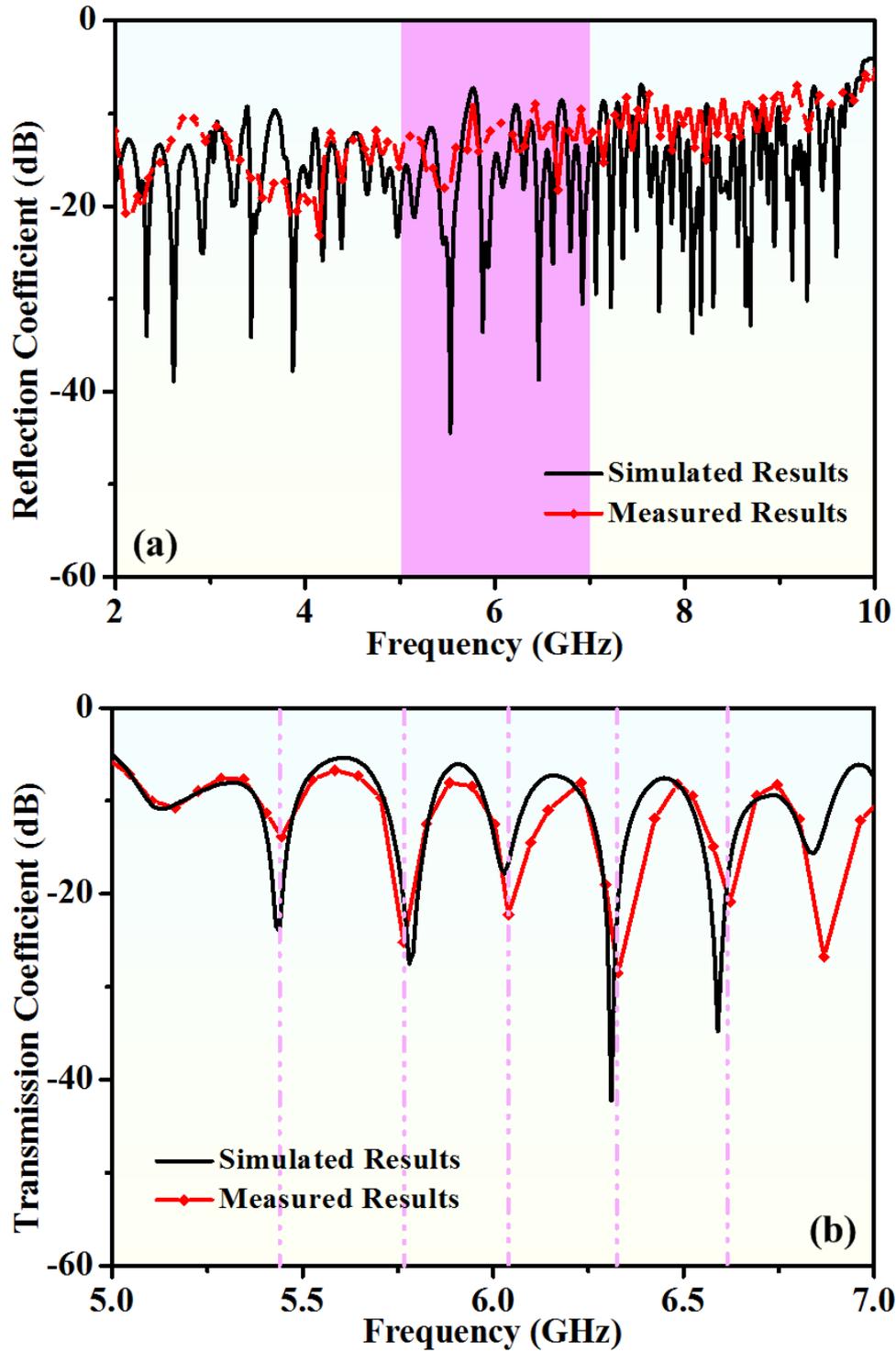

**Figure 4.** Simulated and measured *S* parameters of the proposed structure. (a) Reflection coefficients $S_{11}$. (b) Transmission coefficients $S_{21}$. Short dashed vertical lines from left to right indicate the frequencies of 5.45, 5.78, 6.03, 6.31, 6.59 GHz, corresponding to the OAM mode numbers of *l* =-2, -1, 0, 1, 2.

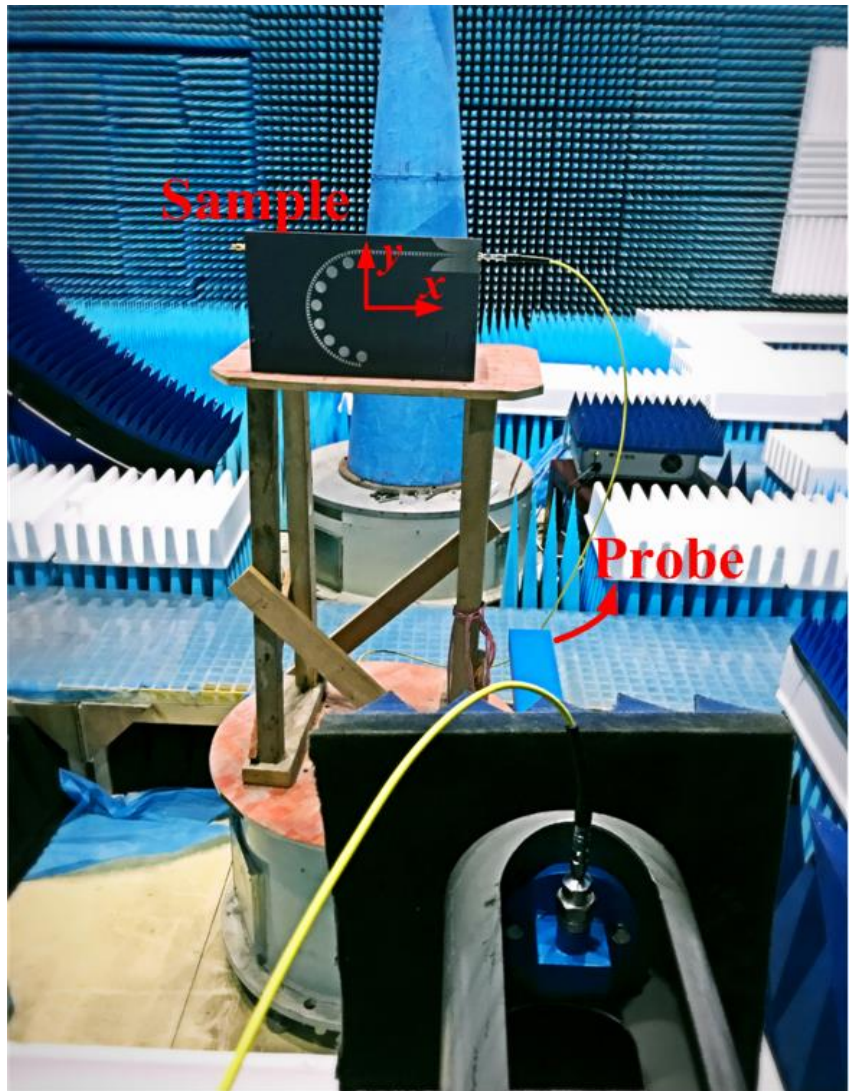

**Figure 5.** Experimental platform in an anechoic chamber.

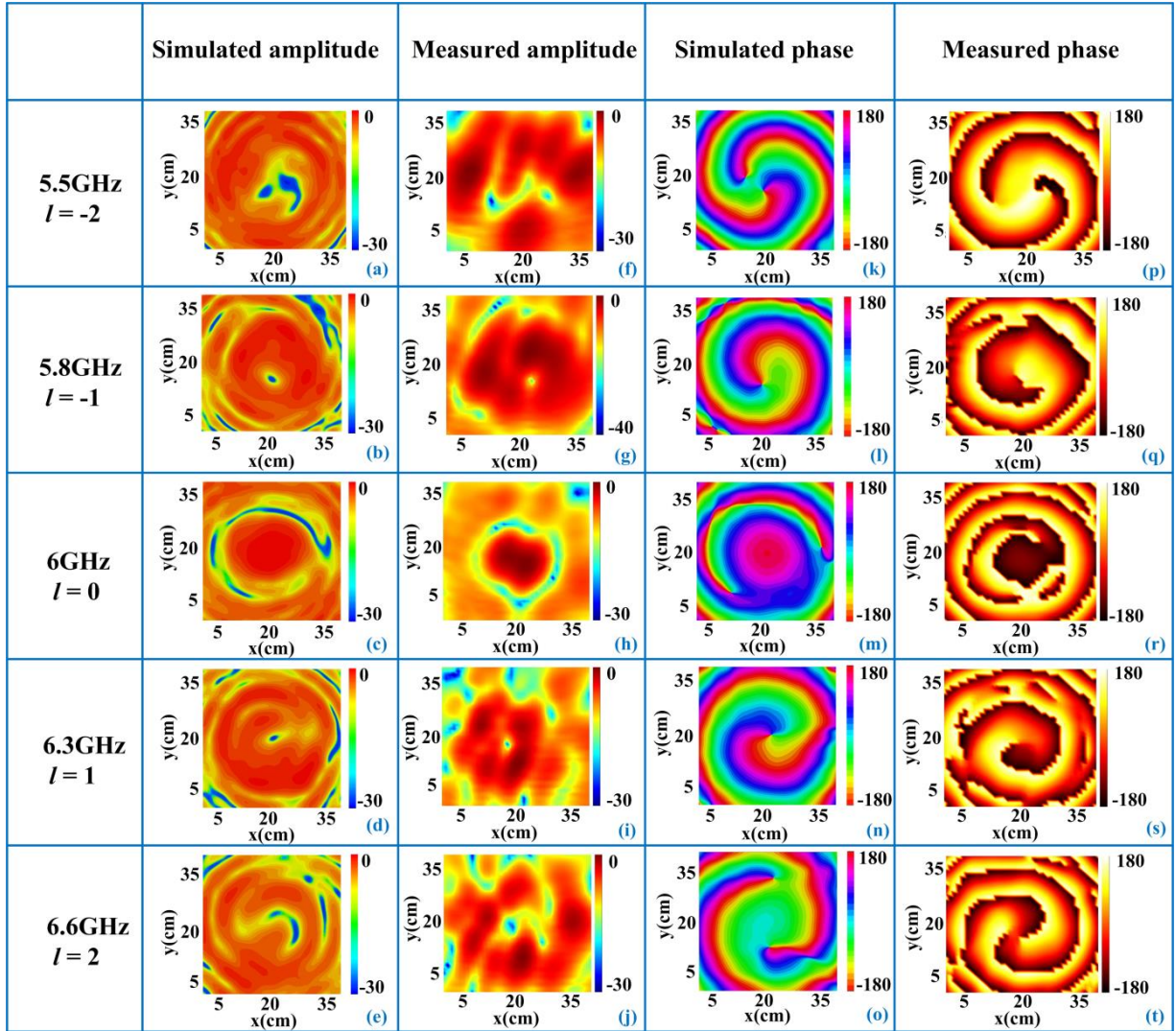

**Figure 6.** Simulated and measured near-field distributions of the OAM modes under different orders.

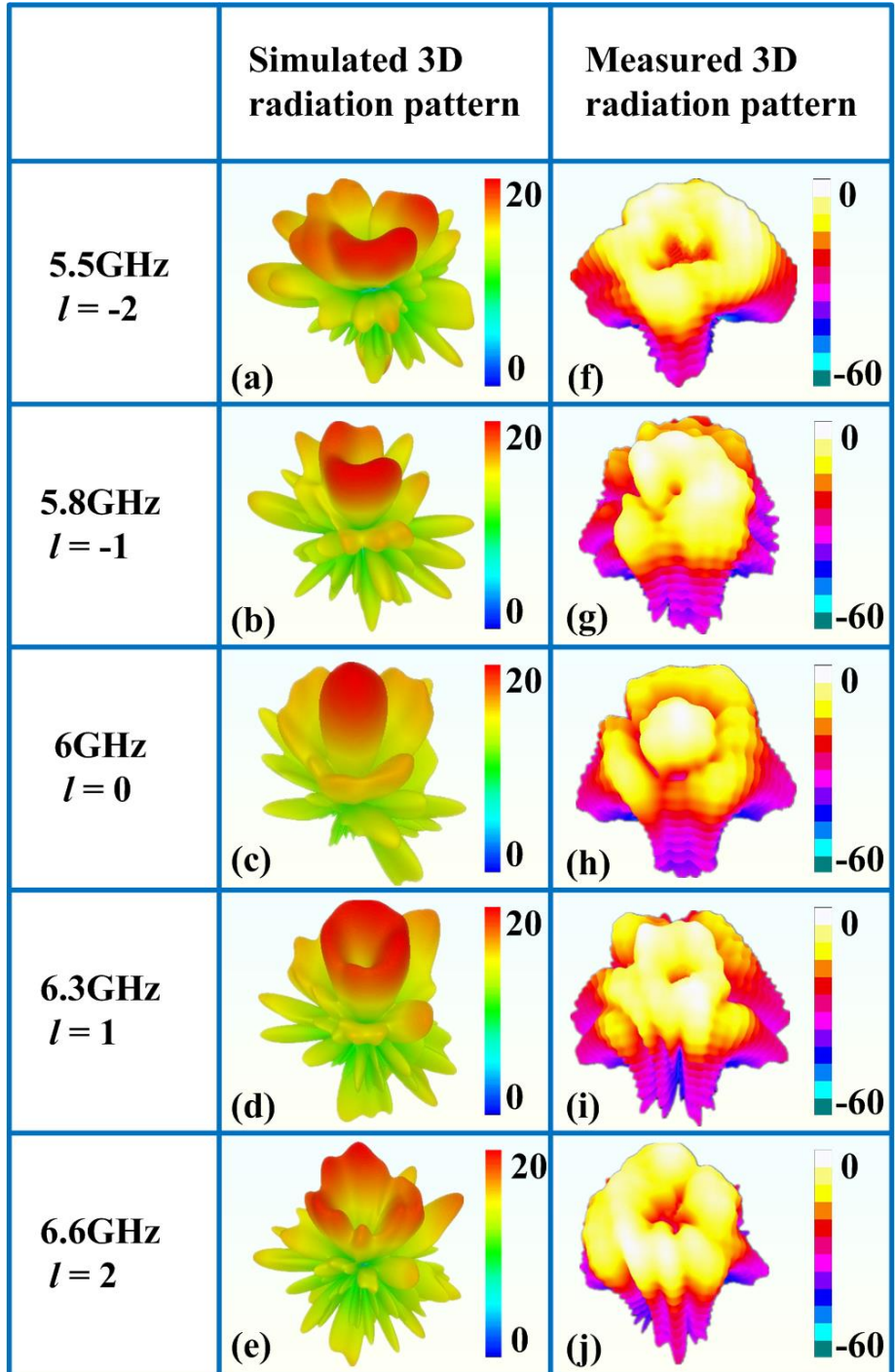

**Figure 7.** Simulated and measured far-field radiation patterns of the OAM modes under different orders.

**Table 1.** Comparison between theoretical arithmetic results and full-wave simulation results of the relevant frequency for each OAM mode.

| Index of OAM mode | Calculated frequency (GHz) (Theoretical arithmetic) | Calculated frequency (GHz) (Full-wave simulation) |
|---|---|---|
| $l = -2$ | 5.41 | 5.45 |
| $l = -1$ | 5.71 | 5.78 |
| $l = 0$ | 6.00 | 6.03 |
| $l = 1$ | 6.28 | 6.31 |
| $l = 2$ | 6.54 | 6.59 |

# Supporting Information

**Microwave Vortex-Beam Emitter Based on Spoof Surface Plasmon Polaritons**

*Jia Yuan Yin[1,2], Jian Ren[3], Lei Zhang[1,2], Haijiang Li[4], and Tie Jun Cui[1,2]\**

To make a preliminary verification of the proposed principle and design, the approximate radiation patterns are calculated through MATLAB. Each circular patch is considered as one of the circular array elements. The formula to calculate the radiation pattern can be written as:

$$S(\theta,\varphi) = \sum_{n=1}^{N} I_n e^{j[ka\sin\theta\cos(\varphi-\varphi_n)+\alpha_n]} \qquad (1)$$

where $I_n$ and $\alpha_n$ represent the amplitude and phase of the nth array element, $a$ is the radius of the circular array, $\varphi_n$ signifies the location of the nth array element, $\theta$ and $\varphi$ are used to describe the coordinate in the space. For simplicity, all the elements are considered possess the same amplitude but with a certain phase difference, which is determined by the distance between the adjacent elements as well as the frequency. That is to say the phase difference between the two adjacent elements is related to the transmission along the spoof SPP waveguide. According to Equation (1), the calculated normalized radiation patterns are shown in **Figure S1**. It is obviously that the normal beam is obtained at 6 GHz, while hollow radiation patterns appear at other frequencies. It is worth noting that, this is only an approximate prediction of the radiation patterns. There is no practical meaning of the numerical values in the figure, since this is just a qualitative analysis.

Gain and efficiency are important indicators of the performance of a beam emitter. With the reference to the simulated results shown in TABLE S1, it can be observed that the proposed design has relatively high radiation efficiency of around 90% generally. And we can see the gain becomes larger from 5.45 GHz to 6.03 GHz, and decreases afterwards. This is owing to the shape properties of vortex beam. When the hollow radiation patterns appear, the radiation energy is scattered instead of being converged to one point. So the gain reaches peak value at around 6 GHz.

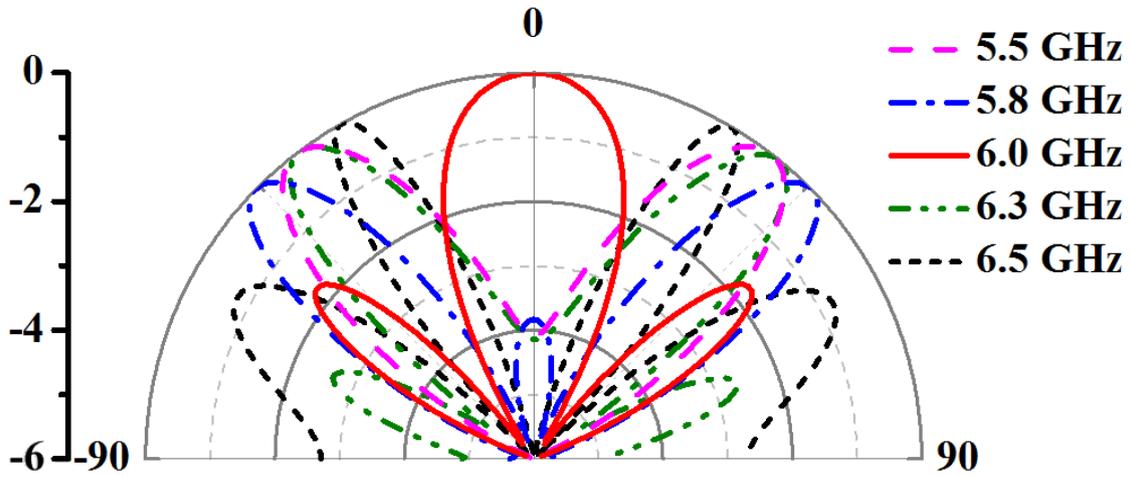

**Figure S1.** Calculated radiation patterns at different frequencies through MATLAB.

**TABLE S1** Simulated gain and efficiency of the proposed structure at relevant frequencies.

| Frequency (GHz) | Gain (dBi) | Efficiency |
|---|---|---|
| 5.45 | 14.39 | 82.94% |
| 5.78 | 15.52 | 89.25% |
| 6.03 | 18.58 | 93.04% |
| 6.31 | 15.86 | 87.87% |
| 6.59 | 15.24 | 89.28% |